# Selective growth of epitaxial $Sr_2IrO_4$ by controlling plume dimensions in pulsed laser deposition

S. S. A. Seo,[1] J. Nichols,[1] J. Hwang,[2] J. Terzic,[1] J. H. Gruenewald,[1] M. Souri,[1] J. Thompson,[1] J. G. Connell,[1] and G. Cao[1]

*[1] Department of Physics and Astronomy, University of Kentucky, Lexington, KY 40506, USA*

*[2] Department of Materials Science and Engineering, The Ohio State University, Columbus, OH 43212, USA*

## ABSTRACT

We report that epitaxial $Sr_2IrO_4$ thin-films can be selectively grown using pulsed laser deposition (PLD). Due to the competition between the Ruddlesden-Popper phases of strontium iridates ($Sr_{n+1}Ir_nO_{3n+1}$), conventional PLD methods often result in mixed phases of $Sr_2IrO_4$ (n = 1), $Sr_3Ir_2O_7$ (n = 2), and $SrIrO_3$ (n = ∞). We have discovered that reduced PLD plume dimensions and slow deposition rates are the key for stabilizing pure $Sr_2IrO_4$ phase thin-films, identified by real-time *in-situ* monitoring of their optical spectra. The slow film deposition results in a thermodynamically stable $TiO_2$\\SrO\$IrO_2$\SrO\SrO configuration at an interface rather than $TiO_2$\\SrO\SrO\$IrO_2$\SrO between a $TiO_2$-teminated $SrTiO_3$ substrate and a $Sr_2IrO_4$ thin film, which is consistent with other layered oxides grown by molecular beam epitaxy. Our approach provides an effective method for using PLD to achieve pure phase thin-films of layered materials that are susceptible to several energetically competing phases.

With the advent of physical vapor deposition techniques such as pulsed laser deposition (PLD), complex oxide thin-films have been extensively studied in applied physics research. Thin-film research offers not only controlled tunability over lattice-symmetry (by strain), dimensionality, and interfacial coupling, but also an important advance for device applications. In particular, PLD has been very successful in synthesizing complex oxide thin-films such as high-$T_c$ superconducting cuprates, e.g., $YBa_2CuO_{7-\delta}$ (Ref. 1). Since PLD is a thermodynamically non-equilibrium process[2], some impurities and defects are inevitably introduced during the growth of a thin film. While these impurities and defects can play a positive role in doping holes or electrons, which are instrumental for inducing superconducting phases, synthesizing a pure phase thin-film via PLD is considered challenging. For example, a layered complex iridium oxide, e.g. $Sr_2IrO_4$, has recently attracted substantial attention for its unique formation of a $J_{eff}$ = 1/2 Mott state from the coexistence of strong spin-orbit coupling and electron-correlation of 5*d* electrons.[3] Recently, unconventional superconductivity is predicted theoretically to be induced in electron doped $Sr_2IrO_4$ (Ref. 4,5), and a *d*-wave gap is observed experimentally[6]. In order to investigate the underlying physical mechanisms further and pave a way for electronic device applications, many research groups have initiated thin-film research on this compound. However, only a few groups have reported the synthesis of $Sr_2IrO_4$ thin films[7-11] compared to the widely successful growth of simple orthorhombic $SrIrO_3$ thin films.

In this letter, we report that epitaxial thin-films of $Sr_2IrO_4$ phase can be selectively grown among the Ruddlesden-Popper (R-P) ($Sr_{n+1}Ir_nO_{3n+1}$) series by pulsed laser deposition (PLD) with *real-time* monitoring of *in-situ* optical spectroscopic ellipsometry. Despite the use of a $Sr_2IrO_4$ polycrystal as a PLD target (i.e., source material), epitaxial thin-films often exhibit various phases even under nearly identical growth conditions. Figure 1 (a) shows the powder x-ray diffraction (XRD) of our $Sr_2IrO_4$ polycrystalline target, which indicates a pure phase of $Sr_2IrO_4$ with no secondary or mixed phases. When the plume size is reduced (Fig. 1 (b1)) by using a smaller laser spot size of approximately 0.25 mm$^2$, we have observed that high-quality $Sr_2IrO_4$ thin-films are grown, as evident in the XRD scan presented in Fig. 1 (b2). However, conventional PLD (with a laser spot size of about 0.75 mm$^2$ and a large plume, as shown in Fig. 1 (c1)) often produces $SrIrO_3$-phase dominant thin films, as shown in the XRD scan of Fig. 1 (c2). A photograph of the two laser spot sizes is presented in Fig. S1. Note that these samples are not pure $SrIrO_3$ thin films but contain other R-P phases as evident from XRD and microscopic characterizations such as transmission electron microscopy (data not shown).

In order to understand this observation systematically, we have synthesized a test sample by using various laser beam spot sizes on the target. (See Ref. 12 regarding the technical details on controlling the laser spot sizes by changing the laser beam aperture. In our PLD system, the focal length of the lens is 50 cm and the distance between the aperture and the lens is about 8 m.) It is somewhat difficult to measure the exact size of the plume. Nevertheless, we have observed that the plume size scales systematically with the laser spot size. While keeping all the growth parameters the same (i.e., a laser (KrF excimer, $\lambda$ = 248 nm) fluence of 1.2 J/cm$^2$, a substrate temperature ($T_s$) of 700 °C, and an oxygen partial pressure ($P_{O2}$) of 10 mTorr), we have discovered that various R-P phases of complex iridates ($Sr_{n+1}Ir_nO_{3n+1}$) such as $Sr_2IrO_4$ (n = 1), $Sr_3Ir_2O_7$ (n = 2), and $SrIrO_3$ (n = $\infty$) are formed depending on the PLD plume dimensions, as shown in the cross-sectional scanning transmission electron microscopy (STEM) image of Fig. 2 (a). In order to prepare a fresh surface for each iridate layer and to distinguish the iridate layers with different laser beam spot sizes, we have inserted $SrTiO_3$ (STO) layers, which is the same material as the substrate. (Atomically flat $TiO_2$-terminated STO substrates are prepared using the method described in Ref. 13.) Figures 2 (b) and 2 (c) show a pure $Sr_2IrO_4$ phase thin-film grown by choosing a laser spot size of approximately 0.25 mm$^2$, without any secondary phases of $SrIrO_3$ and $Sr_3Ir_2O_7$. However, a thin film of pure $SrIrO_3$ or $Sr_3Ir_2O_7$ phase was not obtained presumably because the $Sr_2IrO_4$ polycrystal was used as a target in this work.

Using the *in situ* optical spectroscopic ellipsometry, we have identified the $Sr_2IrO_4$ phase of our thin films in *real time*. The spectroscopic ellipsometry uses a range of photon energies from 1.2 eV to 6.0 eV (1000 nm – 210 nm in wavelength), which allows us to monitor the complex dielectric functions, $\tilde{\varepsilon}(\omega) = \varepsilon_1(\omega) + i\varepsilon_2(\omega)$, of thin films (see Refs. 14,15 for technical details.) Since the optical properties of each phase of the R-P series are quite distinct, the crystallographic phase of the thin film can be determined in *real-time,* which expedites the optimization process and enables the desired $Sr_2IrO_4$ phase to be selectively grown. Figures 3 (a) and 3 (b) show the real dielectric constant ($\varepsilon_1$) and the optical conductivity ($\sigma_1$) spectra of $Sr_2IrO_4$ thin films at the deposition temperature (700 °C) and at room temperature (23 °C), respectively. The complex dielectric constant and the optical conductivity spectra are related by this equation: $\tilde{\varepsilon}(\omega) = \varepsilon_1(\omega) + i\varepsilon_2(\omega) = 1 + 4\pi i \frac{\tilde{\sigma}(\omega)}{\omega}$, where $\omega$ is the photon energy. The high temperature spectra appear overall broader than the room temperature spectra due to the thermal broadening effect. Note that the room temperature optical conductivity spectrum (Fig. 3 (b)) of our $Sr_2IrO_4$ thin film is qualitatively similar to that of a bulk $Sr_2IrO_4$ single-crystal[16]. The peak around 3 eV in the optical conductivity spectra originates from strong charge transfer transitions from the occupied O 2*p* band below the Fermi energy ($E_F$) to the unoccupied Ir 5*d* band above $E_F$. Peaks at low energy below 1.5 eV are characteristic features of the *d*-*d* transitions in the $J_{eff}$ =1/2 and 3/2 bands of $Sr_2IrO_4$ (Ref. 17).

Figure 3 (c) shows the real-time thickness of a $Sr_2IrO_4$ thin film monitored using the *in situ* spectroscopic ellipsometry during PLD. For the real-time thickness monitoring, we have used a single slab model[15] with the thin film thickness as a fitting parameter, as shown in the inset of Fig. 3 (c), and the high temperature dielectric functions of $Sr_2IrO_4$ (Fig. 3 (a) and 3 (b)) and STO (Fig. S2). Note that an extremely slow deposition rate of ca. 300 pulses per 1 nm (i.e. 0.33 Å/sec at a laser pulse rate of 10 Hz) is used in order to synthesize a pure $Sr_2IrO_4$ thin film using a small plume (Fig. 1 (b1)). Recently, Lee *et al.* (Ref. 12) have reported that the kinetic energy of the plume is affected by the laser spot size in PLD so that correct oxygen stoichiometry can be achieved even under oxygen deficient conditions. We speculate that the reduced kinetic energy of the plume with small laser spot size also helps retain correct cation stoichiometry (i.e., correct ratio Sr/Ir = 2) for $Sr_2IrO_4$ during PLD. Since we have kept all other growth parameters fixed in this study, the kinetic conditions seem to play a more important role in stabilizing the $Sr_2IrO_4$ phase with correct cation stoichiometry than thermodynamic conditions.

Figure 4 shows the cross-sectional STEM image and the schematic diagram indicating that the stable arrangement of the interface is $TiO_2$\\SrO\$IrO_2$\SrO\SrO (not $TiO_2$\\SrO\SrO\$IrO_2$\SrO) between a $TiO_2$-terminated STO substrate and a $Sr_2IrO_4$ thin film. According to a recent report on the growth of various layered oxide ($A_2BO_4$) thin films using molecular beam epitaxy (MBE)[18], AO\$BO_2$\AO\AO configuration is known to be thermodynamically stable at a heterointerface since the inversion of layers happens between A-site and B-site ions (i.e, Sr and Ir ions, respectively, in $Sr_2IrO_4$). This observation of MBE growth of layered thin films is consistent with our results from PLD. However, unlike MBE, where separate A-site and B-site ion sources are independently controlled, the PLD technique uses a poly- or single crystalline target so that both A-site and B-site ions are evaporated simultaneously. Further investigations are needed to understand how

layers are inverted occurs during PLD to form a correct SrO\$IrO_2$\SrO\SrO configuration. Nevertheless, the slow growth rate and the reduced kinetic energy of the PLD plume seem to play an important role in stabilizing the initial SrO\$IrO_2$\SrO\SrO layer and subsequent growth of $Sr_2IrO_4$ thin films.

According to a recent paper[19], where a stoichiometric $SrIrO_3$ polycrystal is used as a target, thin films of $SrIrO_3$, $Sr_2IrO_4$, and $Sr_3IrO_7$ phases are stabilized by controlling laser fluence, $P_{O2}$, and $T_s$, which are all parameters typically utilized during conventional PLD. Our results indicate that in addition to utilizing the conventional PLD parameters, manipulating the PLD plume dimension, which is commonly assumed to only influence the growth rate, is instrumental in stabilizing the desired phase of $Sr_2IrO_4$ and is likely a viable tuning parameter for other layered complex oxides.

## SUPPLEMENTARY MATERIAL

See supplementary material for a photograph of focused laser beam spot images and *In-situ* optical spectra of a STO substrate at 700 °C and room temperature.

## ACKNOWLEDGEMENTS

We acknowledge the support of National Science Foundation grant DMR-1454200 for sample synthesis and characterizations.

## Figure Captions

**FIG. 1** **(a)** X-ray $\theta$-2$\theta$ scan of the $Sr_2IrO_4$ polycrystal target used in this work. All the peaks are indexed as pure $Sr_2IrO_4$ phase and no other secondary phases are observed. **(b1)** A photograph of a PLD plume which results in the synthesis of a $Sr_2IrO_4$ epitaxial thin-film. The scale bar is 1 inch (2.54 cm) long. **(b2)** X-ray $\theta$-2$\theta$ scan of an epitaxial $Sr_2IrO_4$ thin-film grown on a STO substrate. Only (00*l*) peaks are identified from the film due to its c-axis orientation. The asterisks (*) are the peaks from the STO substrate. **(c1)** A photograph of PLD plume which results in a $SrIrO_3$ dominant thin-film and **(c2)** x-ray $\theta$-2$\theta$ scan of this sample. The asterisks are the peaks from the $(LaAlO_3)_{0.3}(Sr_2AlTaO_6)_{0.7}$ substrate.

**FIG. 2** Cross-sectional Z-contrast STEM data. The brightest spots are Ir atoms; Sr and Ti atoms are faint, and O atoms are faint due to their small atomic (Z)-number. **(a)** Iridate/STO superlattices exhibiting $SrIrO_3$, $Sr_2IrO_4$, and $Sr_3Ir_2O_7$ phases (the sample is grown as a multilayer, where each iridate layer is deposited with a different laser spot size). **(b)** $Sr_2IrO_4$ single-phase thin films grown under optimal growth conditions using a reduced plume. **(c)** A lower magnification image of the $Sr_2IrO_4$ thin film, which shows that no other R-P series iridates are present.

**FIG. 3** *In-situ* optical spectra (i.e., **(a)** real dielectric constant ($\varepsilon_1$) and **(b)** optical conductivity ($\sigma_1$)) of layered iridate $Sr_2IrO_4$ thin-films during growth at 700 °C (red curves) and room temperature (blue curves). The optical spectra at room temperature are measured after cool-down. The room-temperature $\sigma_1$ spectrum of a $Sr_2IrO_4$ bulk crystal[16] is shown for comparison. **(c)** *Real-time* thickness monitoring of a $Sr_2IrO_4$ thin film using the dielectric functions, shown in **(a)**, and a slab model (inset). Two arrows indicate the start and the end of the deposition, respectively. The slope gives the deposition rate (0.33 Å/sec) at a laser frequency of 10 Hz.

**FIG. 4** Enlarged cross-sectional Z-contrast STEM data near the heterointerface between a $Sr_2IrO_4$ thin-film and a $TiO_2$-terminated STO substrate (left). The dashed line indicates the interface between $Sr_2IrO_4$ and STO. The size of the scale bar is 1 nm. The interfacial configuration is observed as $TiO_2$\\SrO\$IrO_2$\SrO\SrO, not $TiO_2$\\SrO\SrO\$IrO_2$\SrO, as shown in the schematic diagram (right).

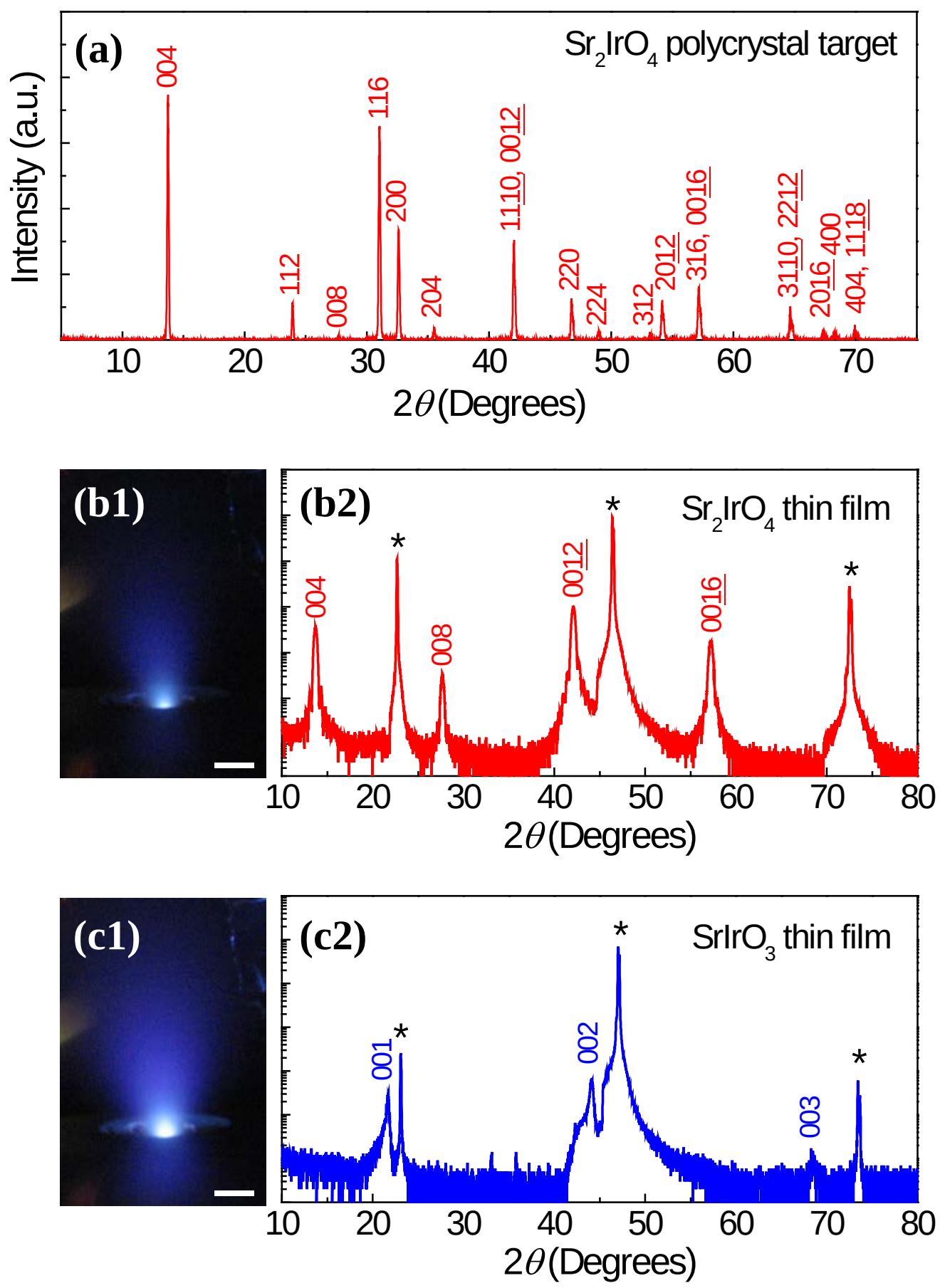

Seo *et al.*
Figure 1

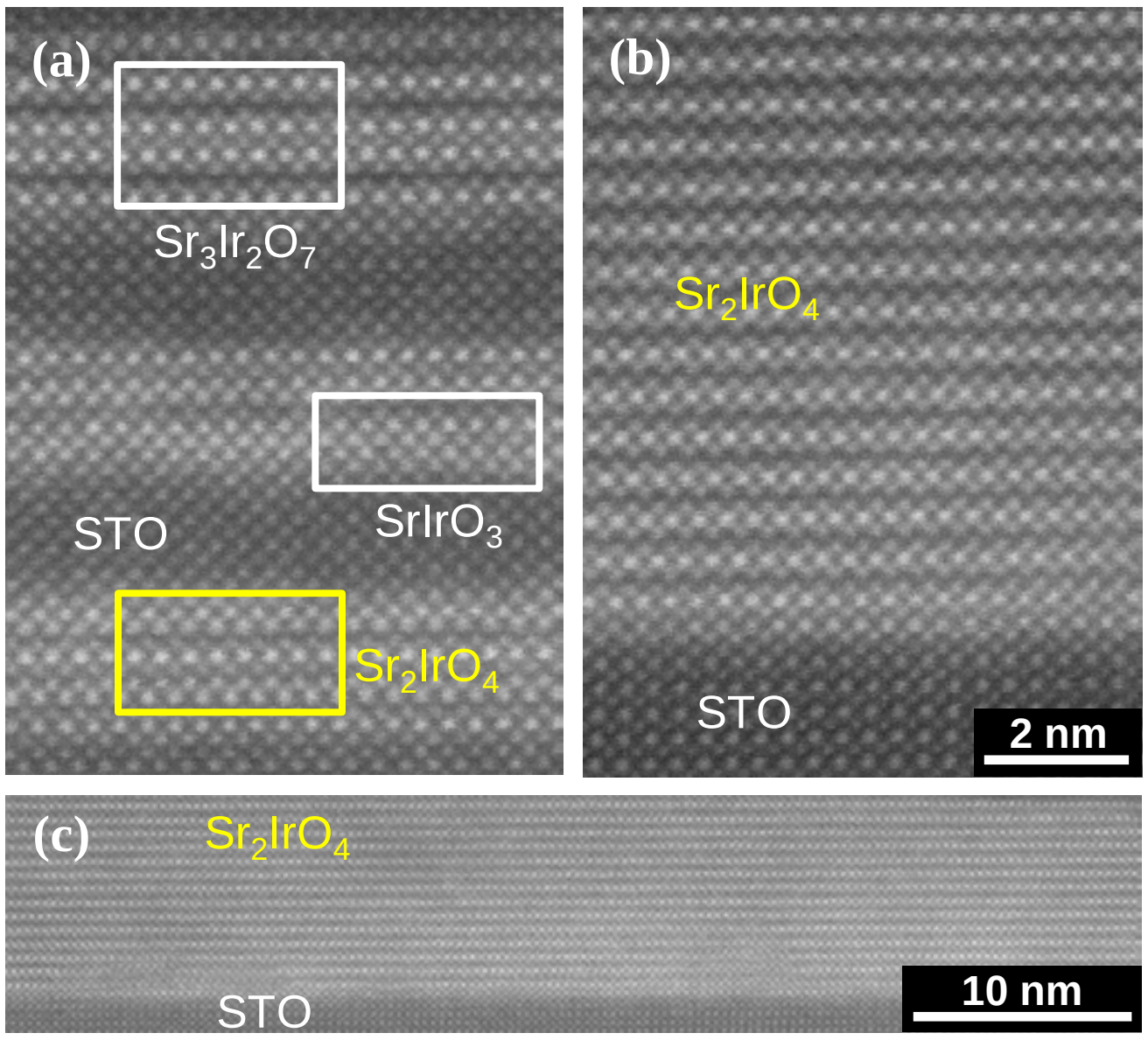

Seo *et al.*
Figure 2

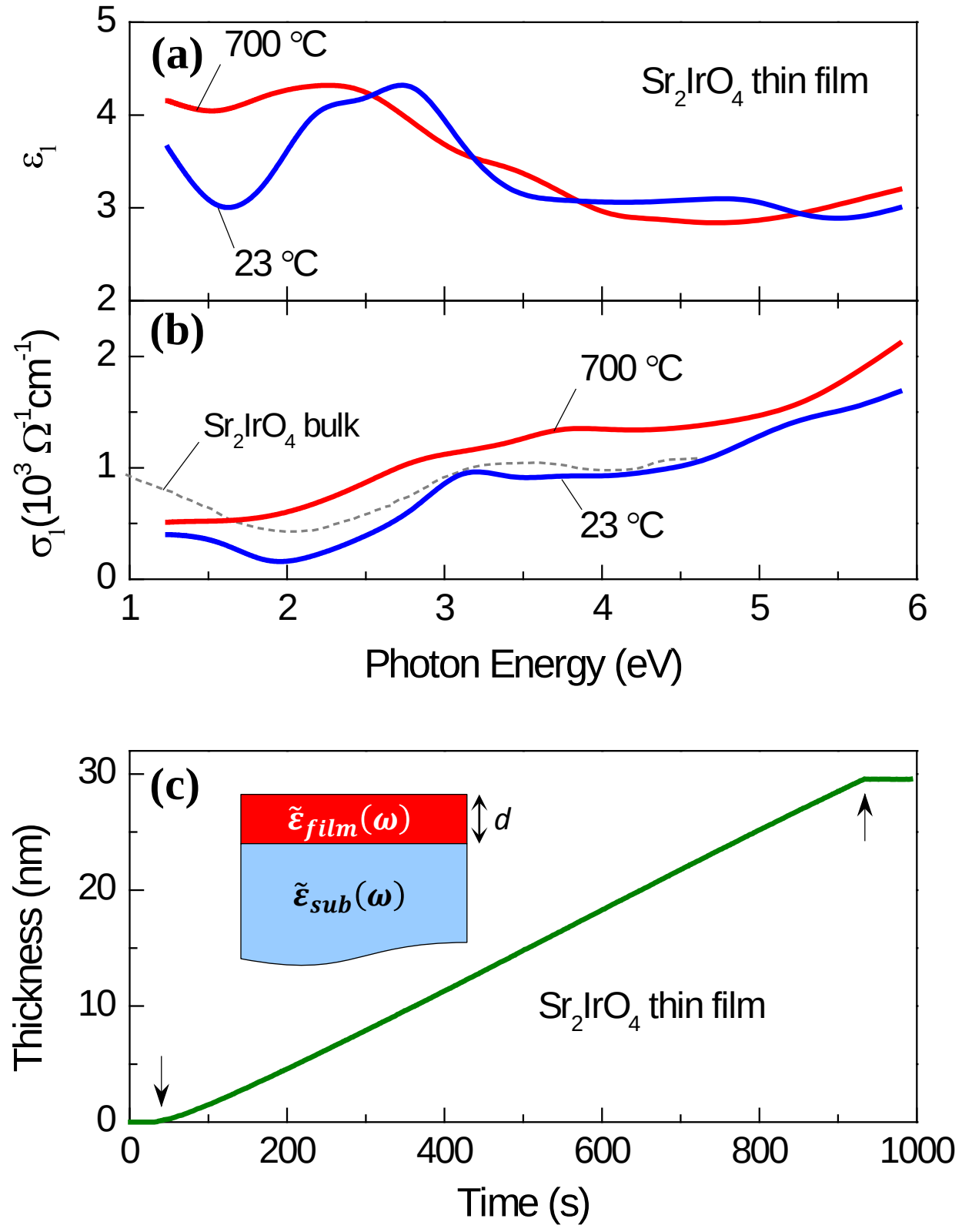

Seo *et al.*
Figure 3

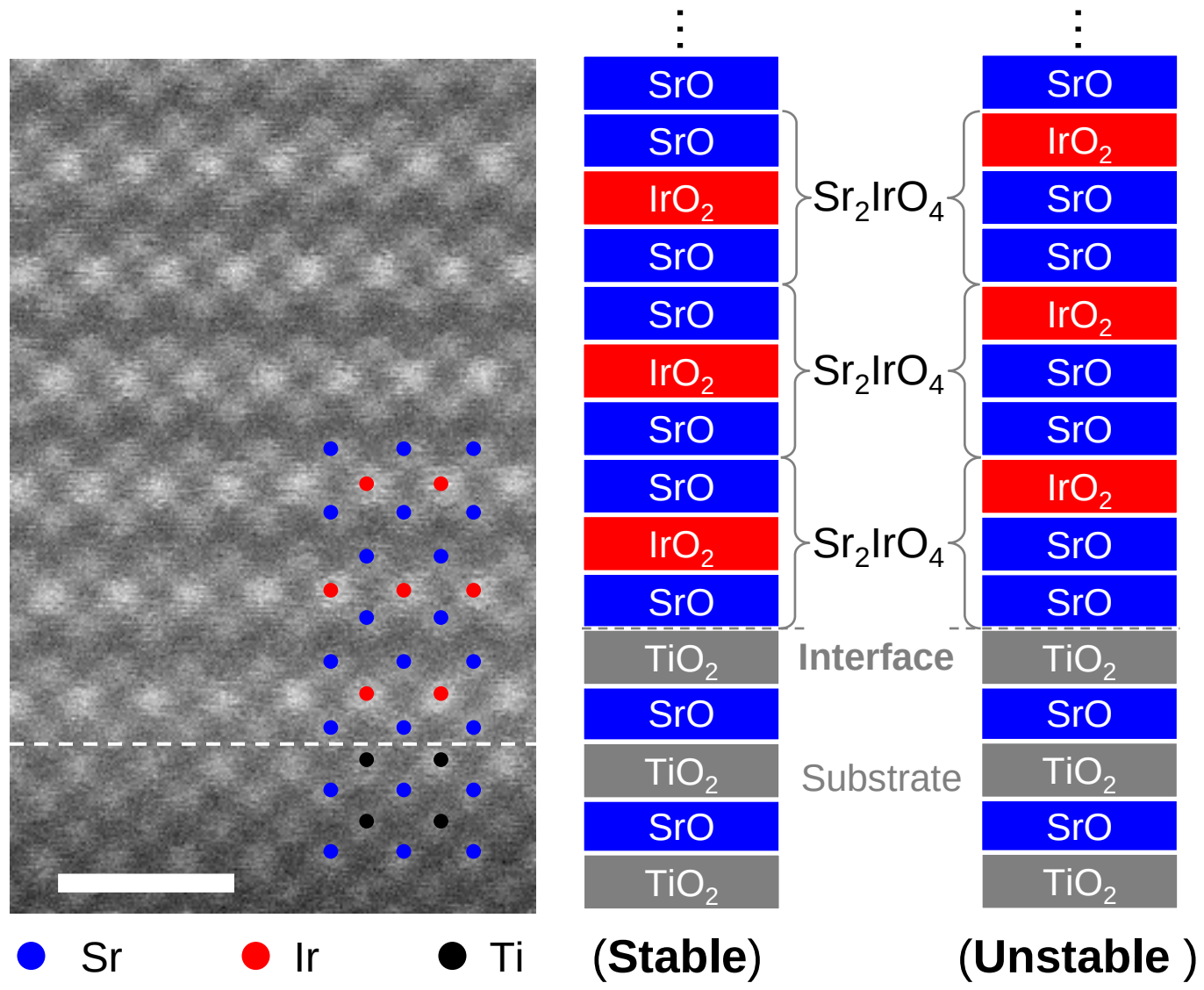

Seo *et al.*
Figure 4

**SUPPLEMENTARY MATERIAL**

# Selective growth of epitaxial $Sr_2IrO_4$ by controlling plume dimensions in pulsed laser deposition

S. S. A. Seo *et al.*

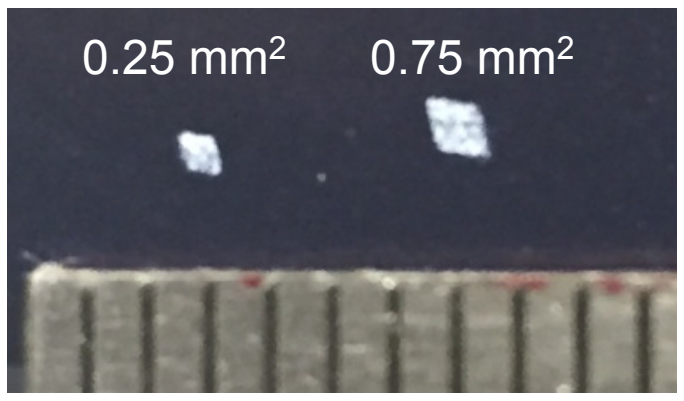

**FIG. S1** A photograph of focused laser beam spot images when plumes of different sizes are generated, as shown in Fig. 1 (b1) and Fig. 1 (c1), respectively.

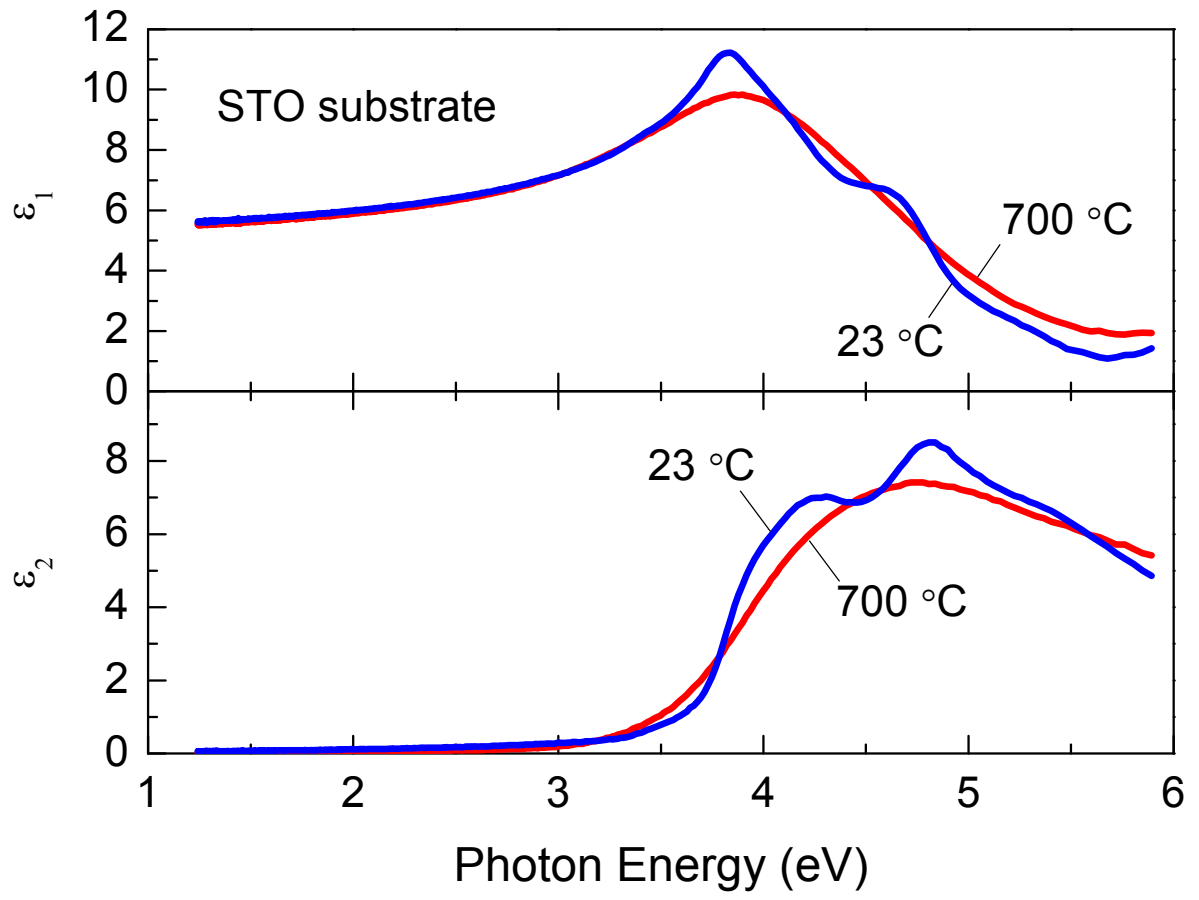

**FIG. S2** *In-situ* dielectric functions of real ($\varepsilon_1$) and imaginary ($\varepsilon_2$) parts of a STO substrate at 700 °C (red curves) and room temperature (blue curves).